# Multi-stable free states of an active particle from coherent memory dynamics

V. Bacot[1,2,+], S. Perrard[1,+,*], M. Labousse[1,2,x], Y. Couder[1,°] and E. Fort[2]

[1]Matière et Systèmes Complexes, CNRS UMR 7057, Université Paris Diderot, Sorbonne Paris Cité, 75013 Paris, France

[2]Institut Langevin, CNRS UMR 7587, ESPCI Paris and PSL University, 75005 Paris, France.

Pressent addresses:

[+]LadHyX, CNRS UMR 7646, École Polytechnique, 91128, Palaiseau, France

[*]Laboratoire de Physique Statistique, CNRS UMR 8550 ENS and PSL University, 75005 Paris, France.

[x]Gulliver, CNRS UMR 7083, ESPCI Paris and PSL University, 75005 Paris, France.

## Abstract

We investigate the dynamics of a deterministic self-propelled particle endowed with coherent memory. We evidence experimentally and numerically that it exhibits several stable free states. The system is composed of a self-propelled drop bouncing on a vibrated liquid driven by the waves it emits at each bounce. This object possesses a propulsion memory resulting from the coherent interference of the waves accumulated along its path. We investigate here the transitory regime of the build-up of the dynamics which leads to velocity modulations. Experiments and numerical simulations enable us to explore unchartered areas of the phase space and reveal the existence of a self-sustained oscillatory regime. Finally, we show the co-existence of several free states. This feature emerges both from the spatio-temporal non-locality of this path memory dynamics as well as the wave nature of the driving mechanism.

The dynamics of a self-propelled particle usually results from the balance between an external friction and a self-propulsive mechanism and leads to a unique ballistic free solution. If the particle interacts with a memoryless thermal bath, a large variety of stochastic paths may be observed [1-3]. However, their dynamics shares common statistical properties. More complex behaviors with several distinct free states are observed in the presence of external sensing, interactions with other particles [4-8] or a minimal form of intelligence [9-10]. In this letter, we show how such complex behaviors emerge even in the absence of any of these mechanisms, from a coherent memory.

We leverage the properties of self-propelled bouncing drops, called *walkers*. Such a system in which the information is stored in waves, was introduced a decade ago [11,12]. This entity is composed of a droplet driven by the surface standing waves it emits when bouncing on the surface of a vertically oscillating bath [13]. The dynamics of this self-propelled object is designated as a path-memory wave driven dynamics. The information about the droplet past trajectory is stored iteratively in its wave field composed of the coherent addition of all elementary standing circular wave fields centered at the successive droplet bounces. In return, this information encoded in the wave drives the droplet dynamics by changing the local slope at each of the impact points. Walkers are endowed with a non-quantum wave-particle duality based on a time non-locality. Several dynamical solutions emerge from the balance between the surface shearing and the propulsive force depending on the excitation parameters which control the memory time. The most natural solution in which the drop moves at a constant speed has been thoroughly investigated [14-18]. Self-trapped spinning states have also been studied [19-22]. Wind-Willassen *et al.* [23] found non-stationary walking states and ascribed these dynamical modes to an alternation between different bouncing modes. Finally, Sampara and Gilet [24] have evidenced that exciting the bath with two frequencies permits non-stationary walking speed. In these two cases, the non-trivial solution arises from the complex bouncing and not from a dynamical feature of the horizontal motion itself.

In this letter, we report the co-existence of free states based on wave interference where the drop vertical motion is synchronized with the bath. We evidence the pivotal role of the memory depth to drive the system into oscillatory or chaotic regimes. We study experimentally and numerically the transient regime of the walk to reveal the process of the memory build-up. We leverage this exploration of the phase space to investigate numerically and experimentally the coexistence of stable non-stationary walking states.

The dynamics of the walkers is successfully described by the path-memory model as soon as the vertical and horizontal motions of the droplet are decoupled [25]. The droplet dynamics is driven by successive kicks that are proportional to the local slope of the global wave field. This wavefield is a coherent superposition of standing waves centered along the droplet trajectory and sustained for a time $\tau$ owing to a critical slowdown at the vicinity of the Faraday instability. This parametric instability originates from the modulation of the apparent gravity at the fluid interface. The decay time of the stationary waves is given by $\tau \propto T_{\text{Far}}(1 - \gamma_m/\gamma_F)^{-1}$ [12,15,16] with $\gamma_m$ the vertical acceleration amplitude of the bath and $\gamma_F$ the Faraday acceleration threshold ($\gamma_F \approx 5$ g in our experiments). The drop hits the bath every Faraday period $T_{\text{Far}} = 40$ ms. The total guiding wavefield $h(\boldsymbol{r}, t_n)$ at position $\boldsymbol{r}$ and time $t_N$ resulting from successive impacts at time $t_k$ and position $\boldsymbol{r}_k$ is

$$h(\boldsymbol{r}, t_N) = h_0 \sum_{k=1}^{N} e^{-\frac{(t_N - t_k)}{\tau}} J_0(k_F |\boldsymbol{r} - \boldsymbol{r}_k|), \qquad (\text{Eq. 1})$$

where $J_n$ denotes the $n$th Bessel function of first kind and $h_0$ is the field amplitude. Each of the previous positions of the drop leaves a wave footprint on the liquid surface and acts as secondary sources. The system possesses a quantifiable amount of "memory" about its past trajectory, to which it is sensitive at all times. The memory parameter $\text{Me} = \frac{\tau}{T_{\text{Far}}}$ measures the mean number of previous bounces contributing to the dynamics of the walker. The walker dynamics is modeled by an iterative numerical scheme that includes a damping force and propelling force proportional to the local slope of the surface field. The algorithm has already been described and benchmarked in Refs [25-28]. If we denote by $v(t_N)$ the 1D horizontal speed of the drop and $x(t_N)$ its position at time $t_N = N\Delta t$, with $\Delta t = T_{\text{Far}}$ then the speed at the next bounce $t_{N+1} = t_N + \Delta t$ is given by

$$v(t_{N+1}) = v(t_N) + \Delta t[-\eta v(t_N) + \mathcal{F}_{\text{wave}}], \qquad (\text{Eq. 2})$$

where the wave force per unit mass is

$$\mathcal{F}_{\text{wave}} = C \sum_{k=1}^{N} \Delta t e^{-\frac{(t_N - t_k)}{\tau}} J_1\big(k_F(x_{N+1} - x_k)\big) \qquad (\text{Eq. 3})$$

and $\eta$ is an effective friction coefficient. We choose the range of parameters $C = 0.2 - 1.5$ m/s$^3$ and $\eta = 2 - 5$ $s^{-1}$ consistently with the hydrodynamic description [15,17]. Note that the exact value crucially depends on the fluid viscosity and drop size [15,17]. Equations (2) and (3) are reminiscent of the integrodifferential equation proposed in Refs [16,19] obtained by taking the continuous time limit $\Delta t \to 0$. Oza *et al*. [16] demonstrated with this continuous time model that the straight line with constant speed is linearly stable to tangential perturbations. Note that this continuous model does not exclude the possibility of other regimes but we find that keeping a finite size step $\Delta t$ is important for the validity of the result presented hereafter (see Supplemental Material Figure S1).

To explore the influence of the dynamical exchange between the wave field and the drop in regimes far from constant velocity steady states, we first focus on the transient regime starting from an immobile drop and increase sharply the bath acceleration. As a result, the memory time increases drastically and the bath starts storing positional information of the drop impacts. Both experiments and numerical simulations have been performed. The experimental set up is sketched in Fig. 1(a). We consider a drop of silicone oil of kinematic viscosity 50 cSt bouncing on a 7 mm deep bath of the same liquid in a square tank of 14 cm long and free of inner obstacles. The tank is vertically vibrated at 50 Hz with an acceleration amplitude gamma $\gamma_m \approx 4.5$g. As shown in Fig. 1(b), we suddenly increase the acceleration amplitude to a value above the walking threshold $\gamma_W$ for which the horizontal motion is observed. Two cameras filming from above and from the side record respectively the horizontal and the vertical motions of the drop.

A snapshot of the experimental transient is shown in Fig. 1(c). The high-speed film reveals how the horizontal motion starts. Immediately after the acceleration increase, the drop bounces a few times on the spot before it starts moving in the horizontal plane. Simultaneously the amplitude of the wave-field increases without changing its form: during the following first bounces, the wave field remains mostly circular, while the drop goes away from the center. As the drop moves forward, the wave-field is deformed and the original bump fully vanishes after the drop has traveled a distance of the order of the wavelength.

The drop positions and speeds are measured and compared with the model [Eq. (2)]. Figure 2(a) shows the time evolution of the horizontal speed for increasing memory parameter Me = 7, 36, and for different drops. At short memory, Me = 7, the drop moves shortly after the acceleration amplitude of the bath $\gamma_m$ is set above the walking acceleration threshold $\gamma_W \simeq 4.8$ g. The time $t_0$ needed to start moving is randomly distributed and depends on the details of the initial conditions. It remained in the range $0.1 - 0.7$ s (*i.e.* 3 to 17 drop bounces) and neither dependence on memory nor on drop size of the average delay have been established from our experiments. For $t_0 \leq t \lesssim 1$ s the drop accelerates sharply and steadily until it reaches a constant velocity $v_f = 6.5 \pm 0.2$ mm/s for $t \gtrsim 1$ s. At a larger memory, Me = 36, the situation changes qualitatively and quantitatively. The drop accelerates more vigorously than for Me = 7 and the speed exhibits oscillations that last for typically few seconds. The speed oscillations are damped in time and the drop speed finally converges to a constant value $v_f = 14.6 \pm 0.1$ mm/s. For large enough memory, typically Me $\gtrsim 20$, the final speed in simulations depends very little on the memory parameter, and mainly on the ratio $C/\eta$ as expected from previous investigations [11,15,16,19].

The temporal variation of drop speed is a consequence of the propelling force modulations which results from the variations of density of secondary sources. To evidence the origin of this dynamical interplay we investigate experimentally the evolution of the frequency of oscillations $\nu$ with Me and with the final drop speed $v_f$. Varying the drop size essentially changes the final drop speed that is an easily measurable quantity. Figure 2(b) shows the experimental evolution of $\nu$ with the final speed of the drop. It reveals that $v_f/\nu$ is constant. The memory parameter does not influence significantly the period of oscillations, so that the characteristic length does not depend on memory. The system presents one single intrinsic relevant length that is the Faraday wavelength of the propelling wave $\lambda_F = 2\pi/k_F = 6.1$ mm. We find $\frac{v_f}{\nu} \approx \lambda_F$ within an uncertainty of $\pm 3 \times 10^{-2}$. We interpret it as an interplay between the dynamical buildup of the wave and the drop oscillations leading to a situation where the motion length scale of the particle $v_f/\nu$ self-adapts with the unique wave lengthscale $\lambda_F$.

This interplay can be analyzed by comparing the instantaneous wave slope with the density of secondary sources. The linear density of secondary sources is given by $\frac{1}{vT_{\text{Far}}}$. We define the relative source density as $\Delta\rho = 1/(vT_{\text{Far}}) - 1/(v_f T_{\text{Far}})$. We compare in Fig. 2 (c) the variation of $\Delta\rho$ with the instantaneous wave force per unit mass $\mathcal{F}_{\text{wave}}$ [see Eq. (3)]. We observe that both quantities oscillate in opposite phases indicating that the dynamical interplay between the drop and its waves originates in a temporal exchange between the horizontal momentum of the drop and the wave force.

We now leverage the existence of non-uniform distribution of secondary sources during the transient to explore the possibility of long-standing non-stationary free walking states. Indeed, transient motion is a practical way to prepare the system in a state where the initial distribution of secondary sources is oscillatory. This situation explores a different region of the phase space where stable speed limit cycles may develop and survive. At this stage, we have only presented situations where the oscillations of the speed were damped in time. We investigate now the existence of parameter regimes leading to permanent speed oscillations. We proceed this systematic exploration with the numerical model before evidencing experimental coexistence of free states. We fix the wave coupling constant $C = 1.1$ m/s$^3$, while varying the damping constant $\eta$ and the memory parameter Me. We measure in the long time limit the amplitude of the fluctuations of speed $\Delta v$ (peak to peak) and construct an order parameter $\chi = \min\{\Delta v/v_f, 1\}$. We show its colormap in Fig. 3(a) in the parameter diagram $(\eta, \text{Me})$. To further characterize these dynamics Fig. 3(b) shows their two-dimensional representation in the phase space $(v, \dot{v})$. The corresponding speed time series are given in the Supplemental Material Figure S2. For low values of $\eta$, the drop moves at a stable constant speed ($\chi = 0$ in Fig. 3(a) corresponding to a fixed point in Fig. 3(b)). For larger damping $\eta$ a finite bandwidth in which the speed oscillations are stable in time, exists ($0 \leq \chi \leq 1$ in Fig. 3(a) corresponding to the stable limit cycle in Fig. 3(b)). Depending on the initial conditions, stable fix points also survive in this regime. The transition $\eta^*(\text{Me})$ between the two regimes is characterized by a strong divergence of the oscillation decay time suggesting a dynamical phase transition. For even stronger damping and large memory, the fluctuations of speed can be of the same order as the final speed. In that critical case the drop stops and turns back which triggers a chaotic regime ($\chi = 1$ in Fig. 3(a) and a strange attractor Fig. 3(b)). Although, chaotic behaviors for synchronous bouncing states have been observed in confined situations [30-33] or in the particular cases mediated by complex bouncing modes [23,24], it is the first time that we observe chaotic *free* regimes of walking droplets which intrinsically rely on their horizontal dynamics. The boundaries separating the three distinct regimes mainly depend on the damping parameter and very little on the memory parameter. A different value of wave coupling, $C$=3.8 m.s$^{-3}$ (supplementary Figure S3), did not qualitatively alter the phase diagram. We expect that the existence of a chaotic free walking regimes itself does not depend on the dimension of the motion but that exact nature of the chaos does [29]. We also note that the transition to chaos is here very different from the chaotic paths observed in confining potentials [30-33]**.**

Figure 3(c) shows the experimental evidence of the two free states: the constant velocity solution and the oscillating velocity solutions (see the corresponding wavefields in the Supplemental Movie SM1). The two free states are attractors and stable. Experimentally, the switch between the two solutions must be triggered by a sufficient external perturbation. For instance, shifting between the two horizontal dynamical modes can occur when the walker reaches the edge of the cell. The initial conditions prescribe which attractor is selected (see numerical simulations in the Supplemental Figure S4). It suggests that this property arises here from a horizontal dynamical interplay between the kinetics of the drop and the energy stored in the wave in contrast with a switch of bouncing modes as in [23]. Indeed, this oscillating velocity mode was observed in [23]. They observe a change of the bouncing vertical dynamics of the droplet associated with a change of the walker speed. Our model predicts this oscillating regime even without taking into account the vertical dynamics. Its role and its contribution to the velocity modulation is still to be investigated.

In this letter we investigate the existence of spontaneous non-trivial free states in a wave memory driven dynamics. A walker is a self-propelled particle which stores information in a wave and it rereads it at a later time. This dynamical exchange of energy between the wave and the particle encodes a rich dynamics which can lead to stable speed oscillations. In contrast with the usual stick-slip, this motion is not associated with frictional variations but is rather driven by oscillations of the propelling force which result from the variations of density of secondary sources. The existence of multiple coexisting free states when the memory strength is further increased is evidenced, which is a striking feature for a system lackings any form of intelligence.

## Acknowledgement

The authors are grateful to M. Hubert for insightful discussions and to L. Rhéa, D. Charalampous and A. Lantheaume for technical assistance. The authors acknowledge the support of the AXA research fund and the French National Research Agency Project FREEFLOW, LABEX WIFI (Laboratory of Excellence ANR-10-LABX-24) within the French Program 'Investments for the Future' under reference ANR-10-IDEX-0001-02 PSL*.

## Captions

**Figure 1** Experimental setup and walker's starting up (a) Schematics of the experimental setup. A bath of silicon oil is vertically vibrated at 50 Hz with a tunable acceleration amplitude $\gamma_m$. It is set to $\gamma_m = 4.5g$ to allow a sub-millimetric drop of the same liquid to bounce in a period doubling regime. (b) Time evolution of the bath acceleration. For $t < 0$, $\gamma_m = 4.5g$ the bouncing drop has no horizontal motion. At time $t = 0$, $\gamma_m$ is suddenly increased above the walking threshold $\gamma_W \approx 4.8$ g for which the drop self-propels on the bath surface, typically $\gamma_W < \gamma_m < \gamma_F \approx 5g$. (c) Snapshots of the starting up of the walker taken with high-speed camera (here time step is $\Delta t = 320$ ms with an image width ≈45 mm). The wave profile is initially circular. The motion breaks this symmetry and accumulating secondary sources changes the form of the wave profile.

**Figure 2:** Starting up characterization of the walker (a) Time evolution of the drop speed for short (Me = 7 (red)) and long memory (blue line (Me = 36)). The two cases correspond physically to two different drops. Experimental data (dashed lines) fitted by the model Eq. 2 (solid lines) with fitting parameters $C = 0.25 \text{ m/s}^3$, $\eta = 3.12 \text{ s}^{-1}$ and $C = 1.1 \text{ m/s}^3$, $\eta = 4.72 \text{ s}^{-1}$ respectively. The walker

converges to the final constant speed $v_f = 6.5 \pm 0.2$ mm/s and $v_f = 14.6 \pm 0.1$ mm/s respectively. (b) Experimental frequency ν of the damped speed oscillation as a function of the final speed for various drops sizes and linear fit (slope 0.164. pink area: 95% confidence interval). (c) Time evolution of the experimental density of secondary sources $\Delta \rho$ along the trajectory (measured in $0.1 \times$ mm$^{-1}$, black) and of the force exerted by the waves on the drop $F_{\text{wave}}$ (per unit mass, m/s$^{-2}$, red) deduced from Eq. (3) for Me = 36. The reference $\Delta \rho$=0 is taken in the final constant speed limit.

**Figure 3:** Existence of multiple free states. (a) Numerical color map of the order parameter $\chi = \min\{\Delta v/v_{\text{f}}, 1\}$ for $C = 1.1$ m/s$^3$ with Me and of $\eta$. Starting from an immobile particle, three distinct regimes are observed: constant speed ($\chi = 0$), stable oscillations ($0 < \chi < 1$) and chaotic regime ($\chi = 1$). Depending on the initial conditions, the region of stable oscillations presents also solutions with a constant speed. The circles along the dashed line at $\eta = 6.2$ s$^{-1}$ at Me = 10 (black), Me = 80 (blue) and Me = 400 (red) correspond to the regimes in Fig. 3(b). (b) Two-dimensional representation of the numerical dynamics in the phase space $(v, \dot{v})$. The dynamics converges to either fixed-point (Me = 10 (black)), speed limit cycle (Me = 80 (blue)), or strange attractor (Me = 400 (red)). (c) Experimental evidence of coexisting free states at Me = 45. Speed of a same drop with respect to its relative position: the drop is locked into a limit cycle (orange diamonds). After perturbations (not shown) it converges to a solution with constant speed (blue square). The motion are straight but initial directions are ill-defined.

**Figure 1**

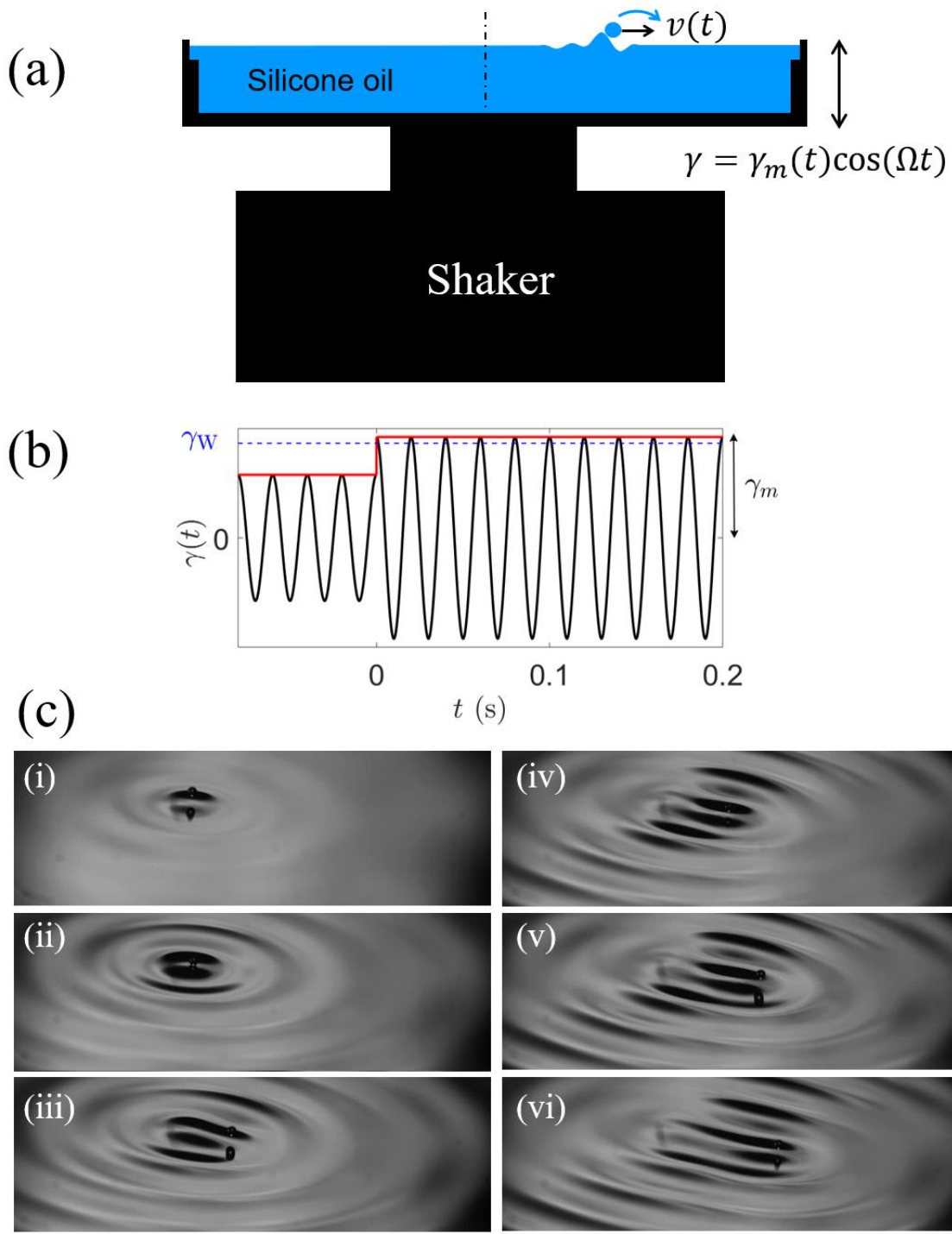

**Figure 2**

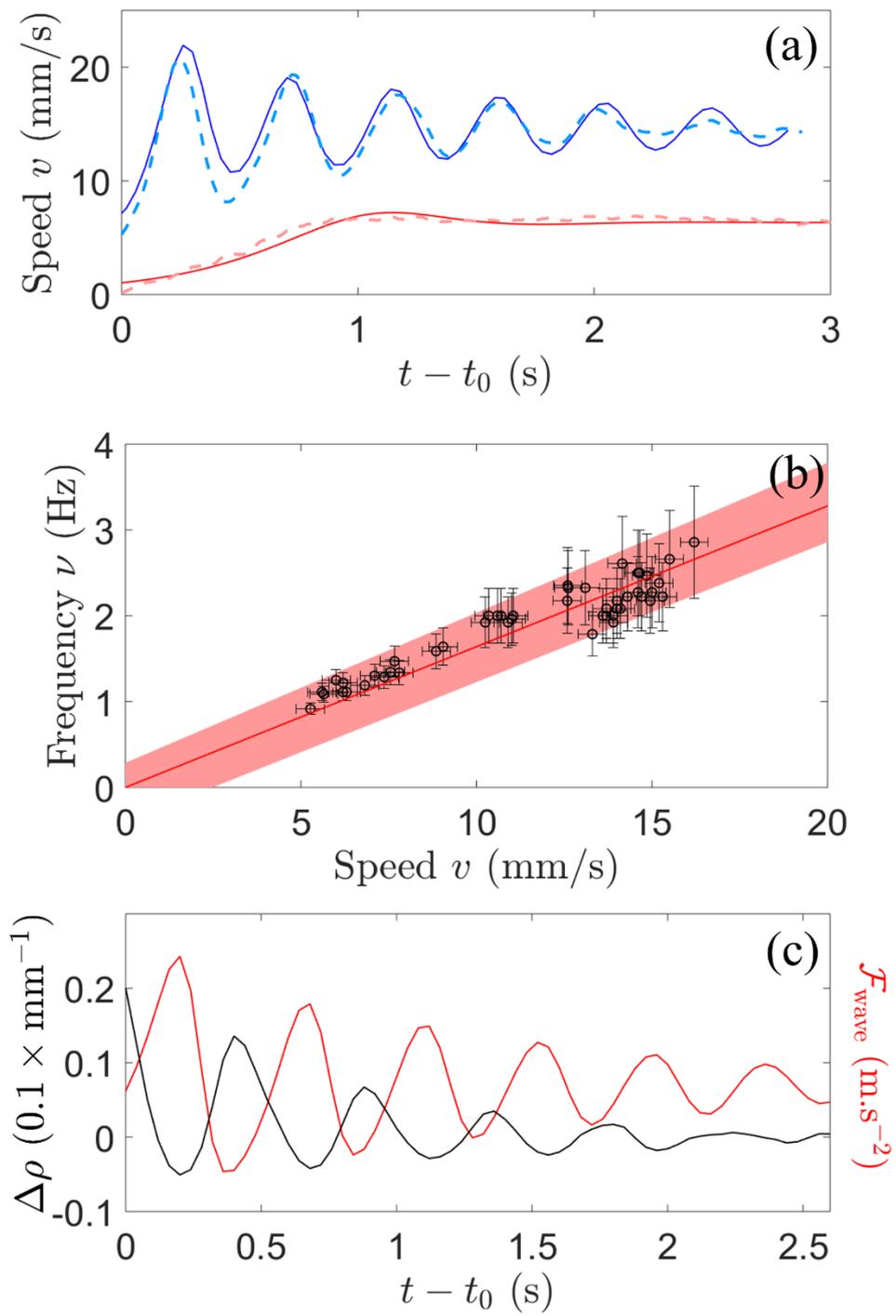

**Figure 3**

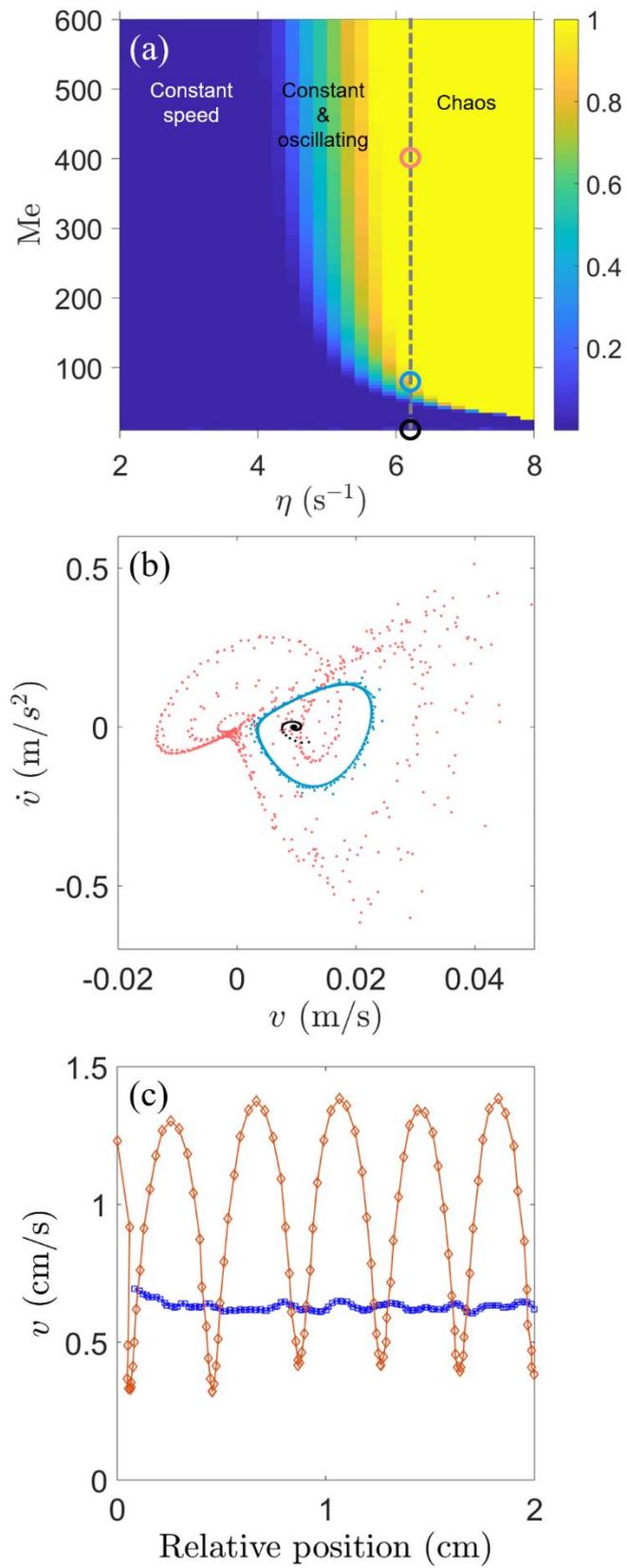